\documentclass[10pt]{article}
\usepackage{amsmath, amsfonts, amsthm, amssymb,verbatim}
        \newtheorem{theorem}{Theorem}[section]
\newtheorem{definition}[theorem]{Definition} 
        \newtheorem{lemma}[theorem]{Lemma}

\numberwithin{equation}{section} 

\newcommand \taut {{\widetilde \tau}} 
\newcommand \Omegah {\widehat \Omega} 
\newcommand \Hessbf {\mbox{\bf Hess \hskip-.06cm}}
\newcommand \Hessbft {\widetilde{\mbox{\bf Hess }}}

\newcommand \taubf      {\mbox{\boldmath$\tau$}}
\newcommand \tautbf      {\mbox{\boldmath$\widetilde\tau$}}

\newcommand \ybf {{\mbox{\boldmath$y$}}}

\newcommand \nablat         {\mbox{\boldmath$\widetilde \nabla$}} 

\newcommand \Mbf {\mathbf M}
\newcommand \pbf {\mathbf p}

\newcommand \qbf {\mathbf q}

\newcommand \Nbf {\mathbf N}

\newcommand \St {{\widetilde S}}

\newcommand \la \langle
\newcommand \ra \rangle
\newcommand \tbar {{\overline t}}
\newcommand \mbar {\overline m}

\newcommand \bart {\underline t}
\newcommand \barm {\underline m}

\newcommand \baru {\underline u}
\newcommand \barc {\underline c}
\newcommand \cbar {\overline c}
\newcommand \ubar {\overline u}
\newcommand \Acal {\mathcal A}
\newcommand \Lcal {\mathcal L}

\newcommand \Fcal {\mathcal F}

\newcommand \bark {\underline k} 
\newcommand \Curv {\mathbf{Curv}}

\newcommand \gbf        {\mathbf g}

\newcommand \Tbf        {\mathbf T}

\newcommand \Dbf      {\mbox{\boldmath$\nabla$}}

\newcommand \Rbf        {\mathbf R}
\newcommand \Bbf        {\mathbf B}

\newcommand \R      {\mathbb R}
\newcommand \be     {\begin{equation}}
\newcommand \ee     {\end{equation}}

\newcommand \kbar   {\overline k}
\newcommand \del        \partial
\newcommand \eps     \epsilon
\newcommand \auth   \textsc

\newcommand \Mcal    {{\mathcal M}}

\newcommand \Jcal   {{\mathcal J}}

\newcommand \Hcal   {{\mathcal H}}
\newcommand \Bcal   {{\mathcal B}}

\newcommand \gt     {{\widetilde {\mathbf g}}}
\newcommand \dt     {{\widetilde {\mathbf d}}}

\newcommand \inj        {\text {\bf Inj}}

\newcommand \Riem       {\text{\bf Rm}}

\newcommand \Rm       {\text{\bf Rm}}
\newcommand \Ric       {\text{Ric}}
\newcommand \Ricbf       {\text{\bf Ric}}

\newcommand \expb   {{{\text{\bf exp}}}}

\begin{document}

\title{Local canonical foliations of Lorentzian manifolds with bounded curvature}
\author{
%                correct spelling :   LeFLOCH  or  LeFloch
Philippe G. LeFloch$^1$}

\date{December 10, 2008}

\maketitle

\footnotetext[1]
{Laboratoire Jacques-Louis Lions \& Centre National de la Recherche Scientifique (CNRS),
Universit\'e Pierre et Marie Curie (Paris 6), 4 Place Jussieu,  75252 Paris, France.
\\
E-mail : {\sl pgLeFloch@gmail.com}  
This work was partially supported by the Agence Nationale de la Recherche (ANR) through the
grant 06-2-134423 entitled {\sl ``Mathematical Methods in General Relativity''} (MATH-GR). 
To appear in: {\it Proc. Workshop on ``Geometry, Topology, QFT, and Cosmology'', May 2008, 
Paris-Meudon Observatory, C. Barbachoux, F. H\'elein, J. Kouneiher, and V. Roubtsov ed.  
}}

\begin{abstract}
We consider pointed Lorentzian manifolds and construct ``canonical'' foliations by constant mean curvature
(CMC)  hypersurfaces. 
Our result assumes a uniform bound 
on the local sup-norm of the curvature of the manifold and on its local injectivity radius, 
only. The prescribed curvature problem under consideration is a nonlinear elliptic equation 
whose coefficients have limited regularity. The CMC foliation allows us to introduce
CMC-harmonic coordinates, in which the coefficients of the Lorentzian metric 
have optimal regularity. 
\end{abstract}

%===========================================================================================

\newpage 

\section{Introduction}
\label{INT}

Spacetimes in general relativity are represented by Lorentzian manifolds satisfying Einstein's field equations.  
For the physical interpretation of the solutions to the Einstein equations, it is important to 
have some insights on their geometrical properties and regularity. To this aim, 
in the present paper we derive 
certain {\sl quantitative estimates} which imply
a sharp control on the local geometry of Lorentzian manifolds. 
More precisely, within a neighborhood of a given observer  
we construct here a {\sl local canonical foliation}  
made of spacelike hypersurfaces whose mean curvature is constant (CMC hypersurfaces) 
and whose geometry is uniformly controled in terms of the curvature and the injectivity radius of the manifold, only.

The proposed method leads to the construction of an ``optimal frame''
 in which the metric coefficients have the best possible regularity 
allowed the sup-norm curvature assumption. 
We derive here new 
geometric estimates that hold under a {\sl limited regularity} assumption on the spacetime, 
while standard techniques apply to more regular spacetimes by 
requiring bounds on first (or even higher) derivatives of the curvature.    
 
The following presentation is based on the papers \cite{ChenLeFloch1,ChenLeFloch2} 
(in collaboration with B.-L. Chen), to which we refer for further applications not covered in the present notes
(especially the construction of CMC-harmonic coordinates, reviewed in \cite{LeFloch}). 
We built here on earlier work on related problems  
by, on one hand, Bartnik and Simon \cite{BartnikSimon} who 
constructed CMC hypersurfaces in the Minkowski spacetime  
and, on the other hand, 
Gerhardt \cite{Gerhardt,Gerhardt1} who established the existence of global CMC foliations in sufficiently 
regular Lorentzian manifolds.

Other earlier related works concern the Einstein equations for vacuum spacetimes. 
One key contribution is Andersson and Moncrief's 
construction of solutions to the Einstein equations 
using CMC-harmonic coordinates  \cite{Andersson0,Andersson,AnderssonMoncrief1}.
Their method allowed them to establish 
a global existence theorem for sufficiently small perturbation of Friedmann-Robertson-Walker type
spacetimes \cite{AnderssonMoncrief2}.   
Therein, first-order derivatives of the 
curvature tensor are assumed to be squared-integrable, at least, 
which is a stronger regularity assumption than the one assumed in Theorem~\ref{main}, below.  

On the other hand, in a series of pioneering papers (\cite{KR1,KR2,KR4,KR5} and the references cited therein), 
Klainerman and Rodnianski initiated an ambitious program about the construction of 
low regularity solutions to the Einstein equations, 
their main results including conjugate radius and injectivity radius estimates, 
which solely involve estimates on the Bel-Robinson energy of the spacetime, that is,
a uniform $L^2$ bound on the curvature tensor on spacelike hypersurfaces.  
Therein, the authors are primarily interested in 
controling the geometry of null cones (i.e.~lights cones) and develop techniques
of analysis for nonlinear wave equations. 
(In contrast, the present paper relies exclusively on elliptic estimates.)   
Still, another direction of research on CMC foliations 
is currently developed by Reiris \cite{Reiris}, who analyzes global aspects of a CMC-Einstein flow 
by requiring higher regularity of the metric.

%============================================================================================================

\section{Local foliation defined by an observer}
\label{BB-01} 

We begin by introducing some notation and then stating our main result (in Theorem~\ref{main}, below). 
For background on Lorentzian geometry we refer the reader to \cite{HawkingEllis,ONeil,Penrose}.
Throughout this paper, $(\Mbf,\gbf)$ denotes a time-oriented, $(n+1)$-dimensional Lorentzian manifold
(with boundary) 
whose Levi-Civita connection and Riemann curvature are denoted by $\Dbf$ and $\Rm$, respectively. 
By convention, the signature of the metric is $(-,+, \ldots, +)$, 
and we recall
 that a tangent vector $X$ is said to be timelike, null, or spacelike if and only if 
 its Lorentzian norm 
$g(X,X)$ is negative, zero, or positive, respectively. 
The time-orientation assumed on $\Mbf$ allows us to distinguish between past- or future-timelike vectors; 
in fact, this assumption is not a real restriction since we are only interested in local properties 
of the spacetime.

A point $\pbf \in \Mbf$ being given, we search for a foliation of a neighborhood of 
$\pbf$ by spacelike hypersurfaces
and we impose that 
each hypersurface has constant (scalar) mean curvature 
and that
its geometry is uniformly controled in terms of geometrically natural quantities.  
Specifically, we will need 
the curvature of the manifold near $\pbf$ and the injectivity radius at $\pbf$, only.
Since the metric $\gbf$ is not positive definite, 
a single point is not sufficient for canonically defining a foliation near $\pbf$
and, in fact, instead of a single point 
we prescribe an {\sl observer} consisting of a pair $(\pbf,\Tbf_\pbf)$, where $\Tbf_\pbf$ is a 
unit future-timelike vector, called a {\sl reference vector}, at $\pbf$. 
It will be convenient to refer to $(\Mbf, \gbf, \pbf,\Tbf_\pbf)$ as a {\sl pointed Lorentzian manifold.}

In a local frame $(e_\alpha)$ at $\pbf$ where $e_0 = \Tbf_\pbf$ coincides with the reference vector
and $e_j$ ($j=1, \ldots, n$) are spacelike vectors, the Lorentzian metric reads 
$$
\gbf= \gbf_\pbf = - e_0 \otimes e_0 + e_1 \otimes e_1 + \ldots + e_n \otimes e_n. 
$$
Clearly, the vector $\Tbf_\pbf$ induces a (positive-definite) inner product on the tangent space at $\pbf$,
defined by 
$$
\gbf_{\Tbf_\pbf} := e_0 \otimes e_0 + e_1 \otimes e_1 + \ldots + e_n \otimes e_n. 
$$
To simplify the notation, we often write $\Tbf=\Tbf_\pbf$ and 
$\la \, \cdot \, , \, \cdot \, \ra_\Tbf := \gbf_{\Tbf_\pbf}$, 
while $|A|_\Tbf = |A|_{\gbf_\Tbf}$ denotes the Riemannian norm of a tensor $A$. 
The corresponding Riemannian ball of radius $r>0$
(as a subset of the tangent space at $\pbf$) is denoted by $B_{\Tbf,r}(\pbf)$.  
In case a vector field $T$ is given, the above construction can be made at each point in a neighborhood
of $\pbf$, and we can define a metric $\gbf_\Tbf$, referred to as the {\sl reference Riemannian metric}
associated with the vector field $\Tbf$.

To state our main result we generalize classical notions from Riemannian geometry, and 
now define the injectivity radius and local curvature bound of an observer in a Lorentzian manifold, 
as follows. 

\begin{enumerate}

\item[] {\bf Injectivity radius of an observer $(\pbf,\Tbf_\pbf)$.}  
Let $\expb_\pbf: B_{\Tbf,r}(\pbf) \to \Mbf$ be the {\sl exponential map} at the point $\pbf$, 
as defined from the Lorentzian metric $\gbf$ 
over the Riemannian ball $B_{\Tbf,r}(\pbf)$. 
This map is well-defined for all sufficiently small radius $r$, at least.
By definition, the {\sl injectivity radius} of the observer $(\pbf,\Tbf_\pbf)$, denoted by 
$$
\inj(\Mbf,\gbf,\pbf,\Tbf_\pbf), 
$$ 
is the supremum among all radii $r>0$ such that the map $\expb_\pbf$ is a {\sl global}
diffeomorphism from $B_{\Tbf,r}(\pbf)$ 
onto its image $\Bcal_{\Tbf,r}(\pbf) \subset \Mbf$.

\item[] {\bf Local curvature of an observer.} By parallel transporting 
(with respect to the Lorentzian connection $\Dbf$)
the given vector $\Tbf_\pbf$ along radial geodesics from $\pbf$, we 
construct a vector field $\Tbf$ defined in a neighborhood of $\pbf$ (at least). 
``Far'' from the base point $\pbf$ this vector field is generally multi-valued, 
since two distinct geodesics leaving from $\pbf$ may intersect.
Given $r>0$ and a radial geodesic $\gamma$ associated with an ``initial'' vector in $\Bbf_{\Tbf,r}(\pbf)$, we 
denote this vector field by $\Tbf_\gamma$ and
we can define 
a positive-definite, inner product $\gbf_{\Tbf_\gamma}$ at each point along $\gamma$. This 
inner-product allows us to compute the 
{\sl local curvature norm of the observer} $(\pbf, \Tbf_\pbf)$ within the ball of radius $r$, defined by
\be
\label{maxcurv}
\Curv(\Mbf,\gbf, \pbf,\Tbf_\pbf; r) := \sup_{\gamma} |\Riem|_{\Tbf_\gamma},
\ee
where the supremum is taken over every radial geodesic from $\pbf$ 
of Riemannian length $r$, at most. 
\end{enumerate}

The above notion of curvature bound
is defined for all $r>0$, even though certain radial geodesics may well be incomplete and 
hit the boundary of $\Mbf$. 
To avoid any difficulty at the boundary we always tacitly assume that all geodesic balls 
under consideration are compactly included in $\Mbf$, so that 
geodesics under consideration can not attain the boundary of the spacetime. 
The notion of curvature bound was originally introduced in \cite{ChenLeFloch1}
where injectivity radius bounds were derived.  
In the situation considered in the following theorem, the radius $r$ under consideration is 
smaller than or equal to the injectivity radius.

Our purpose is to construct a local foliation of a ``large neighborhood'' of the point $\pbf$.

\begin{definition}[Notion of local canonical foliation]
\label{def} 
Fix a parameter $\theta \in (0,1)$. 
Given a pointed Lorentzian manifold $(\Mbf, \gbf,\pbf,\Tbf_\pbf)$, 
a {\rm local canonical foliation associated with the observer} $(\pbf,\Tbf_\pbf)$ is 
any foliation $\bigcup_{\bart \leq t \leq \tbar} \Sigma^t$ containing $\pbf$, 
by spacelike hypersurfaces with constant mean curvature $t \in [\bart, \tbar]$, with 
$$
\bart := (1-\theta) {n \over sr}, \qquad \tbar := (1 + \theta) {n \over sr}, 
\qquad 
s \in [\theta, 2\theta],
$$ 
whose unit normal $\Nbf:=\Dbf t / |\Dbf t|$, 
lapse function $\lambda := (-\gbf(\Dbf t, \Dbf t))^{-1/2}$, and 
second fundamental form $h$ satisfy 
$$
\aligned
1 - g(\Nbf,\Tbf) & \leq \theta^{-1}, \qquad  \theta \leq -r^2 \lambda \leq \theta^{-1},
\\
r \, |h| & \leq \theta^{-1},
\endaligned
$$
in which $\Tbf$ denotes some $\gbf$-parallel translate of $\Tbf_\pbf$ along radial geodesics from $\pbf$. 
\end{definition}

In the above definition, the mean curvature (time variable) is of order $1/(\theta r)$ 
and varies within an interval of length about $\theta/r$. 
For instance, in the four-dimensional Minkowski spacetime, by suitably restricting 
the standard hyperboloidal foliation to a neighborhood of $\pbf$ 
it is easy to determine a canonical foliation: In standard flat coordinates $( \hskip.04cm \widehat t,x^1, x^2, x^3)$ 
with $\pbf= (0, 0,0,0)$ we can find 
$\Sigma^t \subset \big\{ - (  \hskip.06cm \widehat t + c )^2 + (x^1)^2 + (x^2)^2 + (x^3)^2 = - (1/t)^{2} \big\}$
with $c <<1$ so that the estimates in Definition~\ref{def} hold 
and, moreover,  
we can ensure that a Riemannian ball centered at $\pbf$, say 
$( \hskip.04cm \widehat t  \hskip.08cm )^2 + (x^1)^2 + (x^2)^2 + (x^3)^2 \leq \theta^4 r^2$
with $\theta <<1$, is covered by this foliation.

Our main result is as follows. 

\begin{theorem}[Existence of local canonical foliations]
\label{main}
There exists a constant $\theta \in (0,1)$ depending only on the dimension $n$
such that, for every $(n+1)$-dimensional pointed Lorentzian manifold $(\Mbf, \gbf,\pbf,\Tbf_\pbf)$
satisfying the following curvature and injectivity radius bounds at the scale $r>0$
$$
\Curv(\Mbf,\gbf, \pbf,\Tbf_\pbf; r) \leq r^{-2}, 
\qquad 
\inj(\Mbf,\gbf,\pbf,\Tbf_\pbf) \geq r,
$$
there exists a canonical foliation associated with the observer $(\pbf,\Tbf_\pbf)$ that covers 
the reference Riemannian ball $\Bcal_\Tbf(\pbf,\theta^2 r)$, at least. 
\end{theorem}

Observe that the above theorem is purely geometric and, in particular, does not involve 
a choice of local coordinates.  
The conditions on the curvature and the injectivity radius are not restrictive since 
they can always be ensured by suitably rescaling the Lorentzian metric.

The rest of this text is devoted to presenting the proof of Theorem~\ref{main}.

From now on, 
we assume that a pointed Lorentzian manifold is given which satisfies the assumption of the theorem. 
We are going to construct the CMC hypersurfaces as {\sl graphs over Lorentzian geodesic spheres.}  
In addition, Riemannian geodesic spheres associated with the reference metric $\gbf_\Tbf$
will be introduced. 
Both of these types of geodesics spheres 
will serve as ``barrier'' for the prescribed curvature equation discussed below; indeed,    
each CMC hypersurface will be pinched between a Lorentzian geodesic slice and a Riemannian one.
In our construction, it will be important that the parameter $s \in [\theta, 2\theta]$
 (see Definition~\ref{def} above) 
be sufficiently small, that is, the mean curvature of the hypersurfaces should 
be sufficient large.   

To establish Theorem~\ref{main}, one of the main technical difficulties 
is dealing with a nonlinear elliptic equation satisfied by 
the level set function representing the hypersurfaces, 
as well as with the equation satisfied by their second fundamental form. 
A uniform gradient estimate will be derived which guarantee that these equations are uniformly elliptic. 
As we will see, the coefficients have rather {\sl limited regularity} since, at the initial stage of the analysis,
they are only measurable and bounded and the Nash-Moser's iteration technique will be necessary to derive
uniform estimates on the geometry of the foliation.

%==============================================================================================================

\section{Lorentzian vs.~Riemannian geodesic foliations}
\label{sec3}

\subsection*{Lorentzian geodesic foliation}

We are going to introduce two foliations, based on the Lorentzian metric and the reference 
Riemannian metric, respectively, and then 
to formulate the prescribed mean curvature problem in geodesic normal coordinates
$\ybf$ covering a neighborhood of $\pbf$. At this stage of the analysis,  
standard comparison arguments for the Hessian of distance functions,  
yields us a uniform control (although far from being the optimal one) 
of the metric coefficients in the coordinates under consideration.  
We will omit all the proofs in the present section and refer to \cite{ChenLeFloch1} for details. 
It is known (in Riemannian geometry, at least) that coordinates based on distance functions 
do not provide the best regularity of the metric coefficients and, accordingly, 
the geodesic coordinates introduced in the present section serve only 
as an intermediate step in order to construct a canonical CMC foliation 
(and, in turn, CMC-harmonic coordinates).

Throughout, $\barc < c < \cbar<1$ and $C, C_1, \ldots$ denote positive constants that need not be the same at each 
occurrence and are chosen such that 
the constants $1/c$ and $C, C_1, \ldots$, $c/\barc$ are sufficiently large
while the ratios $\barc/c$ and $c/\cbar$ are sufficiently close to $1$. 
Most importantly, these constants depend upon the dimension $n$ of the manifold, only.

The foliation by subsets $\Hcal_\tau$ of Lorentzian geodesic spheres is defined as follows.
Let $\gamma=\gamma(\tau)$ be the future-oriented, timelike geodesic containing $\pbf$ parameterized so 
that, at the parameter value $cr$,  
$$
\gamma(cr)= \pbf, \qquad \gamma(cr) = \Tbf_\pbf, 
$$
and consider a new observer in the past of $\pbf$: 
$$
(\qbf,\Tbf_\qbf) := (\gamma(0), \gamma'(0)). 
$$
Here, the constants $c < \cbar$ are assumed to be sufficiently small 
so that the injectivity radius of the exponential map $\expb_\qbf$ (computed for the new observer) is $\cbar r$, 
at least. This is possible since, by the injectivity radius assumption, 
the exponential map from $\pbf$ is a diffeomorphism covering a neighborhood of $\pbf$ 
and that, thanks to the curvature assumption, 
the injectivity radius at any sufficiently nearby point can also be controled uniformly.

Introduce now geodesic normal coordinates $\ybf = (\tau, y^j)$, 
with  
$0 \leq \tau \leq \cbar r$,
determined from future timelike geodesics originating radially from $\qbf$. 
The Lorentzian metric takes the form 
\be
\label{459}
\gbf = - d\tau^2 + \gbf_{ij} \, dy^idy^j, 
\ee
where $\gbf_{ij}$ denotes the coefficients of the induced metric on the slices of the geodesic foliation.
In these coordinates, the base points $\pbf$ and $\qbf$ are identified with $\pbf = (cr,0,\ldots,0)$ 
and $\qbf = (0,0,\ldots,0)$, respectively.
By construction, the following cone-like region determined by the reference Riemann metric 
$$
\aligned 
& \Jcal^+_{\cbar r}(\qbf) := \expb_\qbf\big(J^+_{\cbar r}(\qbf)\big),
\\
& J^+_{\cbar r}(\qbf) := \Big\{ X \in B_{\Tbf_\qbf,\cbar r}(\qbf) \, : \,  \quad 
X \text{ timelike,}  
\quad
{\la \Tbf_\qbf, X \ra_{\Tbf_\qbf} \over \, \, \la X,X \ra_{\Tbf_\qbf}^{1/2}} \geq 1- \cbar \Big\}
\endaligned
$$ 
can be covered by the chosen coordinate chart.

Therefore, introducing any $\barc < c$ we consider the foliation 
$$ 
\pbf \in \Fcal := \bigcup_{\tau \in [\barc r,\cbar r]} \Hcal_\tau
$$
of some neighborhood of $\pbf$ by spacelike hypersurfaces $\Hcal_\tau$
on which $\tau$ remains constant while the induced metric satisfies (as quadratic forms) 
\be
\label{plat}
C^{-1} \delta_{ij} \leq \gbf_{ij} \leq C \, \delta_{ij}
\qquad \text{in } \Fcal. 
\ee

We introduce the notation 
$$
\bark(\tau,r) := \frac{r^{-1} C_1}{\tan \big(\tau \, r^{-1} C_1 \big)},
\qquad
\kbar(\tau,r)
:= \frac{r^{-1} C_1}{\tanh \big(\tau \, r^{-1} C_1\big)},
$$
where $r^{-1}C_1 \geq r^{-1}$ bounds the (square root of the) curvature. 
(The curvature bound here  
is computed from the point $\qbf$ and may be slightly greater than $r^{-2}$, so that 
one may need $C_1 \geq 1$.)  
Observe that both $\bark(\tau,r)$ and $\kbar(\tau,r)$ behave like $1/\tau$ when $\tau \to 0$
and, furthermore,
that since $\tau \in [\barc r,\cbar r]$ is chosen to be a small multiple of $r$, the function 
$\bark$ is well-defined within the range of interest.  

Consider now the Lorentzian distance function $\taubf$ computed from the (new) base point $\qbf$, 
and let $E:= \big( \Dbf \taubf \big)^\perp$ be the orthogonal complement of its gradient.  
Let $A_{ij} := (-\Hessbf \taubf)|_{E, ij}$ 
be the second fundamental form of the slices of the geodesic foliation, and recall that   
\be
\label{AAA}
A_{ij} = -(\Hessbf \taubf)_{ij} = {1 \over 2} \, \del_\tau \gbf_{ij},  
\ee
where $\del_\tau$ denotes the vector field defined in $\Jcal^+_{\cbar r} (\qbf)$ by parallel transporting the 
vector $\Tbf_\qbf$ along radial geodesics from $\qbf$.

\begin{lemma}[Hessian comparison theorem for the Lorentzian foliation] 
\label{Hessi}  
The second fundamental form of the geodesic foliation is comparable to the induced metric:
\be
\label{Hess}
\bark (\tau, r) \, \gbf|_E \leq A \leq \kbar(\tau,r) \, \gbf|_E
\qquad \text{in } \Fcal.  
\ee
\end{lemma} 

This result is classical for Riemannian manifolds (see for instance \cite{Petersen}, p.~44 and 175)
and, in fact, the standard proof extends immediately to Lorentzian manifolds. The main point
is that the Hessian function satisfies a Riccati-type equation 
$$
\del_\tau (\Hessbf \taubf(X,X)) + (\Hessbf \taubf)^2 (X,X) = -\Riem(X,\del_\tau,X,\del_\tau), 
$$
in which the vector field $X$ is parallel-transported along radial geodesics and so 
satisfies $\del_r X = 0$. The functions $\bark$ and $\kbar$ arise when solving the 
Riccati equation with the Riemann curvature term replaced by its lower or upper bounds implied 
by our curvature assumption.

Lemma~\ref{Hessi} provides us with an important property of the geodesic slices. Namely,
by taking the trace of \eqref{Hess}, 
we see that the mean curvature $H_{\Hcal_\tau}$ of a geodesic slice is
approximately $n / \tau$; more precisely,  
\be 
\label{bard0} 
n \, \bark(\tau,r) \leq H_{\Hcal_\tau}\leq n \, \kbar(\tau,r) 
\qquad \text{in } \Fcal.  
\ee

In view of \eqref{AAA}-\eqref{Hess}, we have a control of the first-order time derivative of the induced metric. 
In addition, the second-order time-derivative can be also estimated 
by expressing the curvature tensor in geodesic normal coordinates
and observing that the corresponding lapse function is identically $1$ 
as stated in \eqref{459} (so that its derivatives are trivially estimated, which 
will not be the case for the forthcoming CMC foliation).   

\begin{lemma}[Uniform bounds for the time-derivatives of the induced metric]
In geodesic normal coordinates, the $\tau$-derivatives of $\gbf_{ij}$ are controled
up to second-order:    
\be
\label{normal}
\aligned
& r^{-1} \Big| \del_\tau \gbf_{ij}  \Big|
  + r^{-2} \Big| \del_\tau^2 \gbf_{ij} \Big| \leq C
&& \quad \text{in } \Fcal.  
\endaligned
\ee 
\end{lemma} 

Hence, in the coordinate under consideration we do have 
a uniform $L^\infty$ control of $g_{ij}$, 
$\del_\tau g_{ij}$, $\del^2_\tau g_{ij}$, 
and $A_{ij}$. 
The forthcoming construction will provide a foliation by constant mean curvature hypersurfaces, 
which has the advantage to {\sl also ensure spatial regularity} of the metric coefficients, 
after introducing spatially harmonic coordinates. 

%----------------------------------------------------------------------------------------------------

\subsection*{Riemannian geodesic foliation}

Consider next the reference Riemannian metric associated with the vector field $\del_\tau$
(which was constructed from $\qbf$ rather than from $\pbf$) 
$$
\gt : = d\tau^2 + \gbf_{ij} \, dy^idy^j.
$$
The bound on the Riemann curvature of the Lorentzian metric can be equivalently expressed in terms of the 
new reference metric, that is,  
\be
\label{curv}
|\Rm|_\gt \leq C \, r^{-2} \quad \text{ in } \Fcal.  
\ee
Pick up an arbitrary point on the geodesic $\gamma$ within a small neighborhood of the point $\pbf$, say 
$$
\pbf' :=\gamma(\tau') \quad \text{ for some } \tau' \in (\barc r, \cbar r). 
$$ 

Then, for every $\taut \in [\barc r, \cbar r]$ (the constants $\barc, \barc$ are always assumed to be sufficiently
close to $c$),  
consider the {\sl Riemannian geodesic spheres} 
$\St_\taut(\pbf')$ centered at $\pbf'$ and associated with the Riemannian metric $\gt$. 
Hence, here the parameter $\taut$ measures the distance from $\pbf'$ to the sphere  $\St_\taut(\pbf')$
which
 are the level sets of the Riemannian distance function $\dt$, that is, 
 $\tautbf := \dt(\pbf', \cdot)$. 
For every $\pbf'$ the following (possibly empty) subsets of Riemannian geodesic spheres 
$$
\Acal_\taut(\pbf') := \St_\taut(\pbf') \cap \Fcal, 
\qquad \taut >0.  
$$
determine a foliation by spacelike hypersurfaces, which contains both $\pbf$ and $\pbf'$ in its interior.

Defining  $\widetilde E:={\big(\nablat \tautbf\big)^\perp}$, 
from the expression of the reference metric we find
\be
\Hessbf \tautbf|_{\widetilde E} 
= \Hessbft \tautbf|_{\widetilde E} - 2 \, (\del_\tau \tautbf) \, A,
\ee
where $\nablat$ and $\Hessbft$ are, respectively,
 the covariant derivative and Hessian operators associated with the metric $\gt$. 
Observing that $\big| \del_\tau \tautbf \big| \leq | \nablat \tautbf |_\gt =1$ and $r \, A$ is uniformly bounded, 
we find:

\begin{lemma}[Hessian comparison theorem for the Riemannian foliation]
The (restriction of the) Hessian of the Riemannian distance function is comparable to the induced metric:
\be
\label{hess}
\big(\bark(\tautbf,r) - C r^{-1} \big) \, \gt|_{\widetilde E}
\leq \big( \Hessbf \tautbf\big)|_{\widetilde E}
\leq
\big (\kbar(\tautbf,r) + C r^{-1} \big) \, \gt|_{\widetilde E}
 \quad \text{ in } \Fcal.  
\ee
\end{lemma}

As was already emphasized, the time-distance, here $\tautbf$, remains less than a small multiple of the spatial distance 
$r$, so that the lower bound in \eqref{hess} is indeed positive. By taking the trace of \eqref{hess}, it follows that for 
$\tau \in [\barc r, \cbar r]$ the mean curvature $H_{\Acal_\tau(\pbf')}$
of the Riemann geodesic slice satisfies the inequalities \eqref{bard0}, that is, 
\be 
\label{bard} 
n \, \bark(\taut,r) \leq H_{\Acal_\taut(\pbf')}\leq n \, \kbar(\taut,r) 
 \quad \text{ in } \Fcal.    
\ee

%--------------------------------------------------------------------------------------------

\subsection*{Local formulation of the prescribed curvature problem}

The foliation of interest $\bigcup_t \Sigma^t$, which we will construct, consists 
of hypersurfaces of constant mean curvature $t$, viewed as graphs
$$
\Sigma^t := \left \{ G^t(y):= (u^t(y),y) \right\} 
$$
over a {\sl given} geodesic slice $\Hcal_\tau$ associated with some value $\tau=sr$ of the time-function.
Here, $t$ will describe some interval $[\bart, \tbar]$ of definite size
and the functions $y \mapsto u^t(y)$ will be determined by solving a nonlinear elliptic boundary-value problem. In the following,
we often write $\Sigma = \Sigma^t$, $u=u^t$, and $G=G^t$.

Recall that $\gamma$ is a fixed, future-oriented, timelike curve passing through $\pbf$, with
 $\gamma(cr)=\pbf$. 
Fix some parameter value $s \in [c,2c]$ and consider the following two points in the future of $\pbf$ 
$$
\pbf_{sr} := \gamma(sr), 
\qquad 
\pbf_{sr}' := \gamma(s'r), 
\quad s' := s+s^2.
$$ 
Then, consider the Riemannian slice $\Acal_{ss'r}(\pbf_{sr}')$ centered at the point $\pbf_{sr}'$.  
Its intersection with the Lorentzian slice $\Hcal_{sr}$ (centered at $\qbf$) 
is non-empty (since $ss' = s^2 + s^3 > s^2 = \dt(\pbf_{sr}', \pbf_{sr})$). This suggests to 
introduce $\Omega_{sr} \subset \Hcal_{sr}$ as the set of points which are ``inside'' the boundary 
{\sl defined} by  
$$
\del \Omega_{sr} := \Acal_{ss'r}(\pbf_{sr}') \cap \Hcal_{sr}.
$$ 
By construction, we have $\pbf_{sr} \in \Omega_{sr}$ and 
$$
\Bcal_{sr, s^{5/2} r/2}\big( \pbf_{sr} \big) \subset \Omega_{sr} \subset \Bcal_{sr, 2 s^{5/2} r }\big( \pbf_{sr} \big),
$$
where $\Bcal_{sr, a}(\pbf_{sr}) \subset \Hcal_{sr}$ is the ball of radius $a$
determined by the induced metric $\gbf_{ij}$ on $\Hcal_{sr}$. (The scaling $s^{5/2}$ follows from 
Cauchy-Schwarz inequality, after noting that the distance between the two slices is $s^2$
and the radius of the Riemannian ball equals $s^2 + s^3$.)

We introduce the following range of mean curvature values 
\be
\label{range}
t \in I(s,r) := [n \, \kbar(sr,r), n \, \bark(2s^2r,r)],  
\ee
since this choice will allow us to use the Lorentzian and Riemannian slices as
natural barriers for the elliptic problem. It should be noted that $\kbar(sr,r) \sim 1/(sr)$
and $\bark(2s^2r,r) \sim 1/(2s^2r)$ with $s <<1$, so that the above interval has non-empty interior.

We seek for a spacelike hypersurface whose mean curvature equals $t$ and whose 
boundary coincides with the one of $\Omega_{sr}$. Analytically, by denoting by $\Mcal u$ the mean curvature 
of the graph of the function $u$, we have to solve the following {\sl Dirichlet problem} 
\be
\label{dirichlet}
\aligned
\Mcal u & = t \qquad \text{ in } \Omega_{sr},
\\
u & = sr \quad \,  \, \text{ in } \del \Omega_{sr}. 
\endaligned
\ee  
For all given (and sufficiently small) $s \in [c,2c]$,  
we will show that this problem has a smooth solution for each $t \in I(s,r)$. 
However, in general, this $t$-foliation need not contain the base point $\pbf$
so that, in order for our construction to be complete,  
it will be necessary to determine a value, say $s_0$, such that the foliation constructed 
over $\Omega_{rs_0}$ does contain $\pbf$ in its interior (cf.~the argument in Section~\ref{proofs}, below).  

The problem \eqref{dirichlet} admits a variational formulation based on maximizing an area functional, which however 
will not be needed here. 
 
%=============================================================================================================== 

\section{Quantitative estimates}

Recall that $(\tau, y^j)$ denote normal geodesic coordinates defined from a point in the past of $\pbf$. 
The induced metric $g_{ij} = \gbf(\del_i (u,y), \del_j (u,y))$ on the slice $\Sigma$  and its inverse 
take the form 
$$
g_{ij} = \gbf_{ij} - u_i u_j,
\qquad
g^{ij} = \gbf^{ij} + {\gbf^{ik} \gbf^{jl} u_k u_l \over 1 - \gbf^{ij} u_iu_j}, 
$$
where we have set $u_j := \del_j u$ and we recall that $\gbf^{ij} = \gbf^{ij}(u,\cdot)$. 
The hypersurface $\Sigma$ is Riemannian if and only if
$\gbf^{ij} u_iu_j < 1$, which we assume. 
Denote by $\nabla$ the covariant derivative associated with the metric $g$ on $\Sigma$. 
The future-oriented unit normal to each hypersurface is
$$
\Nbf = - \nu \, (1, \nabla u), 
$$
where we have introduced $\nu=\nu(u,\nabla u) := \sqrt{1 + |\nabla u|^2}$. 

The mean curvature of the CMC slice reads, in intrinsic form, 
$$
\Mcal u :=  \nu^{-1} \left(
\Delta u + {A_j}^j\right),
$$
where $\Delta$ is the Laplace operator of $(\Sigma,g)$ or, equivalently, in local coordinates
$$ 
\aligned
\Mcal u =
& \gbf^{-1/2} \del_i 
\big( \gbf^{1/2} \,  \nu \, \gbf^{ij} \del_j u \big)
 + {1 \over 2}  \big(  \nu \, \gbf^{ik} \gbf^{jl} \del_k u \del_l u   
+ \nu^{-1}  \gbf^{ij} \big) \del_\tau \gbf_{ij}.
\endaligned
$$ 
%-----------------------------------------------------------------------------------------------------

The proof of Theorem~\ref{main} is decomposed into several lemmas, 
for which we recall once more that all constants depend upon the dimension of the manifold, only. 
Moreover, all of the following statements
concern solutions $u$ of \eqref{dirichlet} (or general functions as far as Lemma~\ref{compa} is concerned)
corresponding to a mean curvature $t \in I(s,r)$.  
We also recall that the constants arising the statements must be chosen either sufficiently small or large, 
according to our notation in the beginning of Section~\ref{sec3}. 
 
 Let us point out an important property which follows from our set-up of the 
 prescribed mean curvature problem. 
The linearized mean curvature operator around
a constant mean curvature hypersurface takes the form 
\be
\label{line}
\Lcal\Mcal (\varphi)=\Delta \big( \nu \, \varphi \big)
      - \big( |h|^2 + \Ricbf(\Nbf,\Nbf) \big) \, \nu \, \varphi.
\ee
The coefficient $\nu(\nabla u)$ is bounded away from zero, so that 
this operator is uniformly elliptic and invertible,
 since by our curvature assumption 
$$
|\Ricbf(\Nbf,\Nbf)|\lesssim 1, 
\qquad
|h|^2\geq \frac{H^2}{n}, 
$$
so that $|h|^2$ dominates $\Ricbf(\Nbf,\Nbf)$ when $H$ is sufficiently large or, equivalently in our construction,  
when $s$ is sufficiently small. Hence, the above operator shares with the standard Laplacian
the same positivity properties.

The first lemma below states that CMC hypersurfaces are ordered monotonically, according to their mean curvature. 
It is a consequence of the maximum principle for elliptic operators.

\begin{lemma}[Comparison principle]
\label{compa}
Given two functions $u,w$ satisfying 
$$
\Mcal u\geq \Mcal w
$$
in their domain of definition $D$ and $u \leq w$ along the corresponding boundary, one has
either
$u<w$ in the interior of $D$,
 or else $u \equiv w$. 
 
 In particular,
if $\Mcal u \geq n \, \kbar(\cbar r, r)$ everywhere
and $u \leq \cbar r$ along the boundary, then $u \leq \cbar r$.
Similarly, if $\Mcal u\leq n\, \bark(\barc r, r)$ everywhere and $u\geq \barc r$ along the boundary, then $u\geq \bar c r$.
\end{lemma}

The following lemmas will be established in the following sections.
From Lemma~\ref{compa}, we deduce the following uniform bound.

\begin{lemma}[Boundary gradient estimate]
\label{boundary}
If $u$ is a solution of \eqref{dirichlet} with mean curvature $t\in I(s,r)$, then 
\be
\label{48}
| \nabla u| 
\lesssim 1 \quad \text{ on the boundary } \del \Omega_s.
\ee
\end{lemma}

From now on, we use the notation $A \lesssim B$ whenever 
$A \leq C \, B$ for some constant $C>0$ that only depends upon the dimension of the manifold. 
Then, we derive an inequality satisfied by the Laplacian of the function $u$ 
on the hypersurface $\Sigma$.

\begin{lemma}[Consequence of Weitzenb\"ock's identity]
\label{nouveau} 
If $u$ is a solution of \eqref{dirichlet} with mean curvature $t\in I(s,r)$, then 
\be
\label{key3}
\aligned
\Delta |\nabla u|^2 - 2 |\nabla^2 u|^2
 & \gtrsim \la \nabla u, \nabla \Delta u \ra
      - \big(1+ |\nabla u|^2\big)^3,
\\
|\Delta u | & \lesssim 1+ |\nabla u|^2. 
\endaligned
\ee
\end{lemma}

Observe that the Laplace operator $\Delta$ 
on $\Sigma$ depends on metric coefficients on which, at this stage of the analysis,
we solely have an $L^\infty$ control. 
To derive the global gradient estimate below, which is one of the main difficulties in ensuring
that the prescribed mean curvature 
equation is uniformly elliptic, 
we use Nash-Moser's iteration technique, which 
is adapted to handle elliptic operators with solely measurable and bounded coefficients.

\begin{lemma}[Spacelike nature of the CMC hypersurfaces]
\label{global}
If $u$ is a solution of \eqref{dirichlet} with mean curvature $t\in I(s,r)$, then 
$$
\sup_{\Omega_{sr}} | \nabla u| \lesssim 1. 
$$ 
\end{lemma}

Now, in view of the a priori estimates in Lemmas~\ref{boundary} and \ref{global}, the general 
continuation techniques for nonlinear elliptic operators \cite{GilbargTrudinger} 
applies.

\begin{lemma}[Existence of CMC hypersurfaces]
\label{folia}
Given some data 
$$
s \in [c,2c], \qquad t\in I(s,r), 
$$
the Dirichlet problem \eqref{dirichlet} admits a solution $u: \Omega_{sr} \to [sr, 2s^2r]$ 
corresponding to a uniformly spacelike graph with mean curvature $t$.  
\end{lemma}

Note that, by Lemma~\ref{global}, the induced metric on
$\Sigma$ is uniformly equivalent to the metric $\gbf_{ij}$ on the domain $\Omega_{sr}$ so that
we can use, for instance, Sobolev inequalities on $\Sigma$.
Then, the following estimate is established using (once more) Nash-Moser's iteration technique.  

\begin{lemma}[Second fundamental form]
\label{sec}
The solution constructed in Lemma~\ref{folia} satisfies, 
for every $q' \in \Sigma \setminus \del\Sigma$,  
$$ 
\aligned
& |h(q') | \lesssim {1 \over d(q', \del \Sigma)},
\endaligned
$$
where $d(q',\del \Sigma)$ is the distance to the boundary $\del \Sigma$
measured with the induced metric $g$ on $\Sigma$.
\end{lemma}

We thus have 
$$
\sup_{\Omegah_{sr}} r |h| \leq \theta^{-1},
$$
where $\Omegah_{sr} := \Bcal_{sr, s^{5/2} r/4}( \pbf_s ) \subset  \subset \Omega_s$. 
So, the bound on the second fundamental form holds only in a subset of $\Omega_s$,
whose diameter, however, is also of the order $r$. 

Observe that the upper bound in the above lemma blows-up if the point $q'$ approaches the boundary of the CMC slice,
but, by keeping $q' \in \Omegah_s$, the factor $d(q', \del \Sigma)$ is of order $r$ at least, as required.

\begin{lemma}[Time-derivative of the level-set function]
\label{tidu}
The solution constructed in Lemma~\ref{folia} satisfies, 
$$ 
r^2 \lesssim - \del_t u \lesssim r^2 
 \qquad \text{ in } \Omegah_{sr}.
$$
\end{lemma}

We emphasize that the above technique also yields further uniform bounds valid for 
each CMC hypersurface, especially an integral estimate for 
$\nabla h$ 
as well as a pointwise estimate on the second fundamental form of its boundary.

In Section~\ref{proofs} below, we establish Lemmas~\ref{boundary} and \ref{tidu}
whose proof is comparatively easier and, then, conclude the proof of our main theorem.  
The (more involved) proofs of Lemmas~\ref{nouveau} and \ref{global}
and of Lemma~\ref{sec} will be the subject of the following two sections. 
From now on, without loss of generality we choose the normalization 
$$
r=1. 
$$

%=================================================================================================================

\section{Existence of the CMC foliation}
\label{proofs}

\subsection*{Boundary gradient estimate} 
 
To establish Lemma~\ref{boundary}, we use the maximum principle stated in Lemma~\ref{compa}, 
which allows us to compare together two hypersurfaces. 

Given a parameter $s \in [c,2c]$,   
we consider the Lorentzian geodesic sphere from $\qbf$  (with radius $s$) 
$$
y \in \Omega_s \mapsto (\ubar(y),y) := (s,y)
$$
together with the Riemannian geodesic sphere from $\pbf_s'$ (with radius $ss'$)  
$$
y \in \Omega_s \mapsto (\baru(y),y).
$$
We recall that $\baru \leq \ubar$ and that, by construction, both functions coincide with
$s$ on their boundary $\del\Omega_s$. 

Our main observation is that every solution $u$ of \eqref{dirichlet} with mean curvature 
$t\in I(s,1)$ satisfies  
$$
\baru \leq u \leq \ubar.
$$
Clearly, this is sufficient to conclude with \eqref{48}, since we already know that both graphs $\baru, \ubar$ 
are uniformly spacelike; this is especially true along the boundary. 

We define the following natural upper and lower bounds for the function $u$: 
$$
\mbar = \sup_{y\in \Omega_s}u(y), \qquad
\barm = \sup_{y\in \Omega_s} \dt((u(y),y),\pbf_s')
$$
and we rely on comparison arguments based on Lemma~\ref{compa}. 
By contradiction, 
suppose that $\mbar \neq s \equiv \ubar$, that is, suppose that the graph of $u$ is above the one of $\ubar$. 
Hence, $u$ would achieve its maximum at some interior point $y_0 \in \Omega_s \setminus \del\Omega_s$.
Now, according to \eqref{bard0},  
the geodesic slice $\tau \equiv \mbar$ has mean curvature within $[n \bark(\mbar, 1), n \kbar(\mbar,1)]$, 
while the graph of $u$ has mean curvature within $[n \kbar(s,1), n \bark(2s^2,1)]$.
Moreover, since their are tangent at the contact point $(y_0,\mbar)$, we conclude that at the point
$(y_0,u(y_0))$ at least the mean curvature of $u$ is {\sl less than} 
or equal to that of $\tau\equiv \mbar$.
However, this is a contradiction since, when $\mbar>s$, one has $\kbar(\mbar,1) < \kbar(s,1)$. 
Therefore, we conclude that, in fact, the graph of $u$ is below the one of $\ubar$. 

Second, let us check that the graph of $u$ is above the one of $\baru$. By contraction, assume that
at some point $y_1 \in \Omega_s$ 
$$
\dt((u(y_1),y_1),\pbf_s') = \barm < ss'.
$$
At the point $(u(y_1),y_1)$ of the graph of the CMC slice, let us compare
the mean curvature of that slice to the one of {\sl another} Riemannian sphere, the one which is tangent 
at that point,  
i.e. the Riemannian sphere centered at $\qbf'$ and with radius $\dt((u(y_1),y_1),\pbf_s')=\barm$. 
Namely, using \eqref{bard} we find 
$$
\Mcal u(y_1) \geq n \bark(\barm, 1) > n\bark(ss',1). 
$$
However, by assumption, the CMC slice has mean curvature $\Mcal u \leq n\bark(2s^2,1)$. This is contraction 
since $2s^2 > ss'$.

%-----------------------------------------------------------------------------------------------------------------

\subsection*{Time-derivative of the level-set function}

To establish Lemma~\ref{tidu} we use a distance comparison theorem within the Riemannian slice $\Sigma$. 
Consider the point $\pbf'' = (u(\pbf_s),\pbf_s) \in \Sigma$ ``above'' the point $\pbf_s \in \Hcal_s$, 
and introduce the geodesic distance function $\rho=\rho(\pbf'',\cdot)$
measured with the induced metric on the CMC hypersurface $\Sigma$. 
Since the mean-curvature is constant and the curvature of the spacetime is bounded, 
the Gauss equation implies that the Ricci curvature of the hypersurface 
is bounded, especially from below, i.e. 
$$
R_{ij} \gtrsim - g_{ij} \qquad \text{ on } \Sigma. 
$$
Then, the Laplacian comparison theorem for distance functions tells us that 
$\rho$ is a supersolution for the operator $-\Delta + C/\rho$, with $C \geq 0$ 
$$
\Delta \rho\leq C \rho^{-1} \qquad \text{ on } \Sigma. 
$$
Let $\varphi \geq 0$ be a non-increasing, cut-off function away from the boundary $\del \Sigma$. 
(See Section~\ref{section77}, below, for further details.). 
We differentiate the equation \eqref{dirichlet} with respect to $t$ 
and use the bounds on $h$ given in Lemma~\ref{sec} (the proof of that result being
 independent from the present argument). 
 Recalling that $|h|^2 + \Ricbf(\Nbf,\Nbf) \geq 0$, 
we obtain
\be
\label{ubb}
\big( \Delta - |h|^2 - \Ricbf(\Nbf,\Nbf) \big)
\Big( \nu(\nabla u) \, \del_t u
 + \eps \, \varphi\big(\frac{4\rho}{s^{5/2}}\big) \Big)
\geq
1-\eps C  \geq 0,
\ee
in which the term $1$ in the right-hand side comes from differentiating the (constant) 
mean curvature $t$ and we have chosen 
$\eps := 1 / C$. Then, observing that $u=s$ along the boundary $\del \Sigma$ (hence $\del_t u = 0$), 
the maximum principle tells us that 
$\del_t u \leq - \nu^{-1} \, \eps \varphi$. Consequently,
since $\nu$ is uniformly bounded (the hypersurface is uniformly spacelike),   
we obtain $\del_t u \leq -C_1 <0$ 
on the subset $\widehat \Omega_s = \Bcal_{s, s^{5/2}/4}(\pbf_s)\subset\{\tau=s\}$. 
On the other hand, the sup norm follows easily from the maximum principle applied
directly to $\big( \Delta - |h|^2 - \Ricbf(\Nbf,\Nbf) \big)
\big( \nu \, \del_t u \big) = 1$ with vanishing boundary conditions on $\del\Sigma$.

%------------------------------------------------------------------------------------------

\subsection*{Proof of the main theorem}
 
First of all, Lemmas~\ref{tidu} and \ref{sec} show that the family of CMC hypersurfaces 
constructed over a given geodesic slice 
forms a foliation whose geometry is uniformly controled. Hence, 
to complete the proof of Theorem~\ref{main}, it remains to check that our construction yields 
to a foliation containing a definite neighborhood of the base point $\pbf$.
For arbitrary $s$, the foliation need not pass through the given observer $\pbf=\gamma(c)$ and 
it is necessary to choose the parameter $s$
to ensure that the foliation contains a definite neighborhood of $\pbf$.

It is convenient to associate {\sl one} specific CMC slice to each geodesic slice. Precisely, 
for each $s \in [c,2c]$ let $u^{(s)}$ be the CMC graph 
constructed over the reference domain $\Omega_s\subset \{\tau=s\}$
for the following {\sl particular choice} of mean curvature parameter 
$$
t= 2 \kbar(s,1) \in I(s,1).
$$
(The precise choice is not too important.)  
As already pointed out, the implicit function theorem applies and, in particular, 
it implies that the family of functions $u^{(s)}$ depends continuously upon $s$.  
We are going to determine a value $s_0$ such that the graph of $u^{(s_0)}$ contains 
the base point $\pbf = \gamma(c)$, 
and the desired family of CMC slices should be then taken to be
the one constructed over that geodesic slice $\Hcal_{s_0}$ (and somehow ``centered around'' 
the CMC slice $u^{(s_0)}$).   

Again, we rely on the monotonicity property in Lemma~\ref{compa}. To each $s$ 
we associate another geodesic slice $\tau= \tau(s)$ defined by 
$$
2\kbar(s,1) = \bark(\tau(s),1).  
$$
(so roughly speaking $\tau(s) \sim s/2$). 
Observe that the graph of $u^{(s)}$ has mean curvature $2 \kbar(s,1)$ by construction, 
while the geodesic slice $\Hcal_{\tau(s)}$ has mean curvature within  $[n \bark(\tau(s), 1), n \kbar(\tau(s),1)]$. 
So, by the comparison principle, the graph of $u^{(s)}$ is {\sl above} the geodesic slice $\Hcal_{\tau(s)}$, 
that is, we have $u^{(s)} \geq \tau(s)$. In particular, taking $s=2c$ and noting that 
$\tau(2c) >c$, 
this implies the hypersurface $u^{(2c)}$ is {\sl above} the geodesic slice $\Hcal_c$, 
which of course contains the base point $\pbf$.  That is, we have $u^{(2c)}(\pbf) \geq c$. 

On the other hand, taking $s=c$, we already know that the whole foliation constructed over the slice
$\Hcal_c$ is {\sl below} this slice. That is,  in particular, $u^{(c)}(\pbf) \leq c$. 
Consequently, by continuity there exists $s_0 \in [c,2c]$ such that 
$$
u^{(s_0)}(\pbf) = c, 
$$
which completes our argument.

%============================================================================================================

\section{Spacelike nature of the CMC hypersurfaces}

\subsection*{Consequence of Weitzenb\"ock's identity}

We now give a proof of Lemma~\ref{nouveau}, starting from Weitzenb\"ock identity for the function~$u$
\be
\label{boch} 
\Delta |\nabla u|^2 - 2 \, |\nabla^2 u|^2 
= 2 \, \Ric(\nabla u, \nabla u)
  + 2 \, \la \nabla u, \nabla \Delta u\ra
\ee
and controling the two terms in the right-hand side.  

The Ricci curvature term is deal with as follows, using Gauss formula
\be
\label{G} 
\aligned
R_{ijkl} = \, &
\Rbf_{\alpha\beta\gamma\delta}G^{\alpha}_iG^{\beta}_jG^{\gamma}_kG^{\delta}_{l}
- \big( h_{ik} h_{jl} - h_{il} h_{jk} \big)\\
= \, 
& \textbf{R}_{ijkl}+\textbf{R}_{0jkl}u_i+\textbf{R}_{i0kl}u_j+\textbf{R}_{ij0l}u_k+\textbf{R}_{ijk0}u_l
    + u_iu_k\textbf{R}_{0j0l}
\\
& +u_iu_l\textbf{R}_{0jk0} +u_ju_k\textbf{R}_{i00l}+u_ju_l\textbf{R}_{i0k0}- \big( h_{ik}
h_{jl} - h_{il} h_{jk} \big),
\endaligned
\ee
in which  $G(y) = (u(y),y)$ is a map from $\Omega_s$ to the spacetime, 
and
$$
G_i = G_*( \del_i )
    = u_i \, \del_\tau + \del_i.
$$
Since the spacetime curvature is bounded, it follows 
(after taking the trace with $g_{jl}$, itself being controled by $(1+ |\nabla u|^2)$) that 
$$
\aligned
\Rbf_{\alpha\beta\gamma\delta} G^\alpha_i G^\beta_j G^\gamma_k G^\delta_l g^{jl}
& \gtrsim -  (1+ |\nabla u|^2) \, \gbf_{ik}
\\
& = -  (1+ |\nabla u|^2) \, (g_{ik} + u_i u_k).
\endaligned
$$
Therefore, by taking the trace in \eqref{G} we conclude that 
$$
R_{ik} \gtrsim h_{il}h_{kj}g^{lj}
         - H \, h_{ik} - (1+ |\nabla u|^2)(g_{ik}+u_iu_k),
$$
where $H=t$ is the mean curvature of the slice. Hence, we obtain the following 
lower bound 
\be
\label{290} 
\Ric(\nabla u, \nabla u) \gtrsim  - (1+ |\nabla u|^2)^3.
\ee

We now consider Laplacian  
\be
\label{equa3}
\Delta u = - A_j^j + \nu(\nabla u) \, t. 
\ee
Lemma~\ref{Hessi} shows that the Hessian of the distance function and, therefore, 
the second fundamental form of the geodesic spheres is 
uniformly controled by the metric, hence
$$
|A_j^j| 
\lesssim  \big| g^{ij} \gbf_{ij} \big| 
\lesssim |\nu(\nabla u)|^2. 
$$
We obtain $|\Delta u | \lesssim|\nu(\nabla u)|^2$ and, together with \eqref{boch} and \eqref{290}, 
this completes the derivation of the inequality \eqref{key3}.

%----------------------------------------------------------------------------------------------------

\subsection*{Spacelike nature of the CMC hypersurfaces} 

Let us turn to the proof of Lemma~\ref{global}, using Nash-Moser's technique. 

\

\noindent{\it Step 1.} Our first objective is estimate $\|\nabla u\|_{L^\infty}$ 
in term of $\|\nabla u \|_{L^{p_0}}$ for some finite $p_0$.   
We introduce the function
$$
v=(\nu^2 - k)_+ := (1+|\nabla u|^2 - k)_+,
$$
where $k$ is chosen to be so large that $v$ vanishes on the boundary $\del\Sigma$, which
is possible by Lemma~\ref{boundary}. 
 
Given $q \geq 1$, we proceed by multiplying the first equation in \eqref{key3} by $v^q$ and integrating
over $\Sigma$. After applying Green's formula and using the second inequality in \eqref{key3}, 
we arrive at 
$$
\aligned
& \int_\Sigma \Big( q \, v^{q-1} |\nabla v|^2 + v^q
\, |\nabla^2 u|^2 \Big) \, dv_\Sigma
 \lesssim \int_\Sigma \Big( q \, v^{q-1} \la \nabla v, \nabla u\ra \, f
  + v^{q+3} + v^q \Big) \,  dv_\Sigma.
\endaligned
$$
Setting $q=:2m-1$ and neglecting the (favorable) second-order term $v^q \, |\nabla^2 u|^2 $, 
we obtain for all $m \geq 1$ 
\be
\label{key5}
\| \nabla v^m \|_{L^2(\Sigma)}^2 
\lesssim 
m^2 \, \| v^{2m+2}+v^{2m-2}\|_{L^1(\Sigma)}. 
\ee 
Setting now $p:=2m-1/2$, \eqref{key5} takes the following form in the coordinates $y$:  
\be
\label{key5b}
\int_{\Omega_s} \gbf^{ij} \del_i(v^{p/2})  \del_j(v^{p/2})  \, dy  
\lesssim 
q^2 \int_{\Omega_s} \big( v^{p+2} + v^{p-2} \big)\, dy,
\ee
as this follows by noting 
$$
|\nabla v^m |^2 \, \sqrt{det(g)} 
\gtrsim \gbf^{ij} \del_i(v^{p/2})  \del_j(v^{p/2}) \, \sqrt{det(\gbf)},
$$

Next, applying the following Sobolev's inequality 
(in a fixed compact domain)
$$
\| w \|_{L^{2n/(n-1)}(\Omega_s)}^2  
\lesssim 
\|  \gbf^{ij} \del_i w  \del_j w + w^2  \|_{L^1(\Omega_s)}
$$
to the function $w := v^{p/2}$, we deduce that for all $p>2$
\be
\label{key5c}
\| v \|_{L^{pn/(n-1)}(\Omega_s)}  
\lesssim
 p^{2/p}  
  \| v^{p+2}+v^{p-2} \|_{L^1(\Omega_s)}^{1/p} 
\ee
Roughly speaking, 
this estimate provides a control of the $L^{pn/(n-1)}$ norm of $v$ in terms of its $L^p$ norm. 
Since $pn/(n-1)<p$, an iteration procedure (described now) allows us
to actually control the sup norm of $v$.

Without loss of generality, we may assume that $\| v\|_{L^\infty(\Omega_s)}  \geq 1$, for
otherwise the result is immediate. Then, \eqref{key5c} leads to the main estimate
$$
\aligned
& \max\big(1, \| v \|_{L^{pn/(n-1)}(\Omega_s)} \big)
\lesssim p^{2/p} \| v\|_{L^\infty(\Omega_s)}^{2/p}  \max\big( 1, \| v \|_{L^p(\Omega_s)} \big).
\endaligned
$$
By iteration we arrive at 
$$
\aligned
& \| v\|_{L^\infty(\Omega_s)}
\lesssim
 \| v\|_{L^\infty(\Omega_s)}^\alpha \| v\|_{L^{p_0}(\Omega_s)}
\\
& \alpha:= {2 \over p_0} \sum_{k=0}^\infty (1-1/n)^k = {2n\over p_0}
\endaligned
$$
which, provided $p_0>2n$, implies that the sup norm of $v$ is uniformly
bounded by its $L^{p_0}$ norm.

\

\noindent{\it Step 2.} We now derive a uniform gradient estimate in  fixed $L^{p_0}$ norm.
From \eqref{key3} we have $|\Delta u| \lesssim|\nu(\nabla u)|^2$, for all $\lambda>0$
we deduce that 
$$
\aligned
\Delta (e^{\lambda u}) 
& = \lambda^2 e^{\lambda u} |\nabla u|^2 + \lambda e^{\lambda u} \Delta u 
\\
& \gtrsim \lambda^2 e^{\lambda u} |\nabla u|^2 - \lambda \, e^{\lambda \, u}  |\nu(\nabla u)|^2.
\endaligned
$$ 
From this and the first inequality in \eqref{key3}, we deduce 
$$
\aligned
\Delta \left( v^q e^{\lambda u} \right)
\gtrsim \, 
& - v^{q-1} \, e^{\lambda u} \Big( \nu^2 (\nu^4 + \lambda v) - \lambda^2 (\nu-1)
\Big)  
+ \lambda q v^{q-1} e^{\lambda u} \la \nabla u, \nabla v \ra 
\\
& + q v^{q-1} e^{\lambda u} \la \nabla u, \nabla (\Delta u) \ra 
+ q (q-1) v^{q-2} e^{\lambda u} |\nabla v|^2.
\endaligned
$$
Observe that $\nu^2 (\nu^4 + \lambda v) - \lambda^2 (\nu-1)
\lesssim v^3$, provided $k>1$ is fixed and $\lambda$ is arbitrarily large. 
Integrating over $\Sigma$, 
proceeding as in Step~1, and taking $\lambda$ to be arbitrary large, we arrive at
$$
\int_\Sigma |\nabla u|^q \, dv_\Sigma \lesssim 1, 
$$ 
which completes the proof of Lemma~\ref{global}.

%=====================================================================================================

\section{Estimating the second fundamental form}
\label{section77}

Our proof of Lemma~\ref{sec} relies on 
Simons' identity \cite{Simons} for the second fundamental form of an hypersurface. In our case, 
the mean curvature of the hypersurface $\Sigma$ under consideration 
is constant and, recalling that $\Nbf$ denotes its normal, we have  
$$
\aligned
\Delta h_{ij} 
& = \Delta h_{ij} - (tr h)_{ij}
\\
& = |h|^2 \, h_{ij}-(tr h)
h_{ik}h_{lj}g^{kl}-\Rbf_{ipjq}h_{kl}g^{pk}g^{ql}+\Rbf_{jplq}h_{ik}g^{pq}g^{kl}
\\
& \quad +\nabla_{p}(\Rbf_{qj\Nbf i})g^{pq} -
\nabla_{j}(\Rbf_{i\Nbf}). 
\endaligned
$$
Multiplying this equation by $h^{ij}$, re-ordering the terms, and using the curvature assumption, 
we find 
\be
\label{simon2}
\aligned
& \Delta |h|^2 - |\nabla h|^2 - |h|^4
+
\la - \nabla_p (\Rbf_{qj\Nbf i})g^{pq} + \nabla_j (\Rbf_{i\Nbf}), h_{ij}\ra
\\ 
& \gtrsim \,  - (|h|^3 + |h|^2). 
\endaligned
\ee
in which the divergence-like structure of the terms involving the curvature will be used thereafter.

Our objective is to derive, from \eqref{simon2},
an estimate for $\|h \|_{L^\infty}$
which will become degenerate near the boundary. 
We again apply Nash-Moser's iteration technique, 
whose application now is comparatively more involved than in the previous section, 
since it is now necessary to use a nested family of domains of integration.

Fix a smooth, non-increasing cut-off function $\varphi:[0,\infty) \to \R_+$ 
such that $\varphi(z) =1 $ for $z \in [0,1/2]$ and $\varphi(z)=0$ for $z \in [1,\infty)$. 
Furthermore, we require that 
$$
|\varphi'|^2 \lesssim |\varphi|.
$$ 
Choose also a point $q' \in \Sigma \setminus \del\Sigma$ and consider the distance 
$\delta:=d(q',\del \Sigma)$ to the boundary of the slice. 
Now, the function $\psi: \Sigma \to \R_+$ defined by 
$$
\psi := \varphi \circ \kappa, \qquad 
\kappa := {d(q',\cdot) \over \delta} 
$$
vanishes near the boundary $\del \Sigma$.

Given any $q \in [1,\infty)$, we multiply \eqref{simon2} by $\psi \, |h|^q$
and integrate over the slice $\Sigma$. Then, after using Green formula and the curvature assumption, we find 
$$
\aligned
& \int_\Sigma \psi \, |h|^q \, \Delta |h|^2  \, dv_\Sigma
\\
& \gtrsim \int_\Sigma \Big(
\psi |h|^q \, \big( 2 \, |\nabla h|^2 + 2 \, |h|^4 
      - C (|h|^3 + |h|^2) - (q+1) \, |\nabla h|\big)
       - |\nabla \psi| \, |h|^{q+1} \Big) \, dv_\Sigma.
\endaligned
$$
It is immediate to check the following upper bound for the left-hand side of the above inequality 
$$
\int_\Sigma \psi \, |h|^q \, \Delta |h|^2  \, dv_\Sigma
\leq
\int_\Sigma 2 \, |\nabla \psi| \, |h|^{q+1} \, |\nabla h| \, dv_\Sigma
  - \int_\Sigma 2 q \, \psi \, |h|^q \, |\nabla |h||^2 \, dv_\Sigma 
$$
and, therefore, using Cauchy-Schwartz's inequality 
\be
\label{CS}
\aligned
& \int_\Sigma \psi \, |\nabla h|^2 \, |h|^q + \int_\Sigma \psi \, |h|^{q+4}  \, dv_\Sigma
\\
&
\lesssim
\int_\Sigma \Big( \big( |\varphi'|^2 \delta^{-2} \varphi^{-1} + \varphi \big)
\circ \kappa \, |h|^{q+2}
 + (q+1)^2 \, \psi \, |h|^q
 + \delta^{-1} \, \left| \varphi' \circ \kappa \right|  \, |h|^{q+1} \Big)  \, dv_\Sigma,
\endaligned
\ee
in which we used $|\varphi'|^2 \, |\varphi|^{-1} \lesssim 1$.

First of all, neglecting the term $ \psi \, |\nabla h|^2 \, |h|^q $ in \eqref{CS} 
and 
using H\"older's inequality to estimate the right-hand side of the above inequality, 
we arrive at  a control of the $L^{q+4}$ norm of the second fundamental form: 
\be
\label{iter}
\aligned
\| \psi^{1/(q+4)} \, |h| \|_{L^{q+4}(\Sigma)} 
& \leq C_q \, \delta^{\frac{n}{q+4}-1}.
\endaligned
\ee
Observe that the constant $C_q$ depends on $q$. 

Next, we take advantage of the favorable term $\psi \, |\nabla h|^2 \, |h|^q$ in \eqref{CS}. 
By Lemma \ref{global}, the hypersurface is uniformly spacelike
and, therefore, the Sobolev inequality holds 
$$
\aligned
\| \psi \, |h|^{q+2} \|_{L^{1-1/n}(\Sigma)} 
\lesssim 
\| \nabla(\psi |h|^{q+2})  \|_{L^1(\Sigma)}.
\endaligned
$$
So, from \eqref{CS} and by suitably choosing a sequence of functions $\varphi$, we find
(for all $i=1,2,\ldots$) 
\be
\label{toiter} 
 \aligned
& \| h \|_{L^{\frac{n(q+2)}{n-1}} ( \Bcal(q',\frac{\delta}{2}+\frac{\delta }{2^{i+1}}) )} 
   \leq
\big(2^i\, C' (q+2)^2 \delta^{-1} \big)^{1/(q+2)}
  \| h \|_{L^{q+2} ( \Bcal(q',\frac{\delta}{2}+\frac{\delta }{2^{i}}) )} 
\endaligned
\ee

It remains to iterate \eqref{toiter} and this leads us to 
\be
\label{Moser}
\sup_{\Bcal(q',\delta/2)} |h|
\leq C_q \, \delta^{-\frac{n}{q+2}} \,   \| h \|_{L^{q+2} ( \Bcal(q',3\delta/4) )}.
\ee
Choosing $q=0$ in \eqref{iter} and $q=2$ in \eqref{Moser}, we conclude 
$\sup_{\Bcal(q',\delta/2)} |h| \lesssim 1/\delta$.

%============================================================================================================
 
\section*{Acknowledgements}

The author is grateful to Lars Andersson for bibliographical informations, and  
thanks C. Barbachoux, F. H\'elein, J. Kouneiher, and V. Roubtsov
for their invitation to the Workshop on 
{\em Geometry, Topology, QFT, and Cosmology,} which was hold at the Observatoire de Paris-Meudon
in May 2008.  Part of this paper was written when the author was visiting the Institut Henri Poincar\'e
in the Spring 2008 during the Semester Program ``Ricci Curvature and Ricci Flow''  
organized by G. Besson, J. Lott, and G. Tian. 
It was completed when the author visited the Mittag-Leffler Institute
in the Fall 2008 during the Semester Program ``Geometry, Analysis, and General Relativity''
organized by L. Andersson, P. Chrusciel, H. Ringstr\"om, and R. Schoen.

%============================================================================================================


\addcontentsline{toc}{section}{References}

\begin{thebibliography}{99}
  

\bibitem{Andersson0} \auth{L. Andersson,}
Constant mean curvature foliations of flat space-times,
Comm. Anal. Geom. 10 (2002), 1125--1150.

\bibitem{Andersson} \auth{L. Andersson,}
Bel-Robinson energy and constant mean-curvature foliations,  
Ann. H. Poincar\'e 5 (2004), 235--244.  

\bibitem{AnderssonMoncrief1} \auth{L. Andersson and V. Moncrief,}
Elliptic-hyperbolic systems and the Einstein equations,
Ann. Inst. Henri Poincar\'e 4 (2003), 1--34.

\bibitem{AnderssonMoncrief2} \auth{L. Andersson and V. Moncrief,}
Future complete vacuum spacetimes, 
in ``The Einstein equations and the large scale behavior of gravitational fields'', 
Birkh\"auser, Basel, 2004, pp.~299--330. 

\bibitem{BartnikSimon} \auth{R. Bartnik and L. Simon,}
Spacelike hypersurfaces with prescribed boundary values and mean curvature,
Commun. Math. Phys. 87 (1982), 131--152.
 
\bibitem{ChenLeFloch1} \auth{B.-L. Chen and P.G. LeFloch,}
Injectivity radius estimates for Lorentzian manifolds,
Commun. Math. Phys. 278 (2008), 679--713.

\bibitem{ChenLeFloch2} \auth{B.-L. Chen and P.G. LeFloch,} 
Local foliations and optimal regularity of Einstein spacetimes, 
preprint, 2008. 

\bibitem{Gerhardt} \auth{C. Gerhardt,}
H-surfaces in Lorentzian manifolds, Commun. Math. Phys. 89 (1983), 523--533.

\bibitem{Gerhardt1} \auth{C. Gerhardt,}
{\sl Curvature Problems,} in ``Series in Geometry and Topology'', vol. 39, International 
Press, Somerville, MA 2006.

\bibitem{GilbargTrudinger} \auth{D. Gilbarg and N.S. Trudinger,}
{\sl Elliptic partial differential equations of second order,}
Springer-Verlag, Berlin, 1983.

\bibitem{HawkingEllis} \auth{S. Hawking and G.F. Ellis,}
{\sl The large scale structure of spacetime,}
Cambridge Univ. Press, 1973.

\bibitem{KR1} \auth{S. Klainerman and I. Rodnianski,}
Ricci defects of microlocalized Einstein metrics,
J. Hyperbolic Differ. Equa. 1 (2004), 85--113.

\bibitem{KR2} \auth{S. Klainerman and I. Rodnianski,}
Rough solutions of the Einstein-vacuum equations,
Ann. of Math. 161 (2005), 1143--1193.

\bibitem{KR4} \auth{S. Klainerman and I. Rodnianski,}
On the radius of injectivity of null hypersurfaces,
J. Amer. Math. Soc. 21 (2008), 775--795. 

\bibitem{KR5} \auth{S. Klainerman and I. Rodnianski,}
On the breakdown criterion in general relativity, preprint, 2008. 

\bibitem{LeFloch} \auth{P.G. LeFloch,}
Existence of CMC-harmonic coordinates for spacetimes with bounded curvature,  
Actes S\'emin. Th\'eor. Spectr. G\'eom. (2008). 
 
\bibitem{ONeil} \auth{B. O'Neill,}
{\sl Semi-Riemannian geometry with applications to relativity,}
Academic Press Inc., New York, 1983.

\bibitem{Penrose} \auth{R. Penrose,}
{\sl Techniques of differential topology in relativity,}
CBMS-NSF Region. Conf. Series Appli. Math., Vol.~7, 1972.

\bibitem{Petersen} \auth{P. Petersen,}
{\sl Riemannian geometry,}
Springer Verlag, 2nd edition, 2006. 

\bibitem{Reiris} \auth{M. Reiris,}
The constant mean curvature Einstein flow and the Bel-Robinson energy, Preprint,
 ArXiv:0705.3070.  
 
\bibitem{Simons} \auth{J. Simons,}
Minimal varieties in Riemannian manifolds, 
Ann. of Math. 88 (1968), 62--105.


\end{thebibliography}
\end{document}